\documentclass[prd,preprint,tightenlines,nofootinbib]{revtex4}

\usepackage{axodraw}
\usepackage{color}
\usepackage{amsmath}
\usepackage{multirow}
\usepackage{graphicx}
\usepackage[tight]{subfigure}

\newcommand{\Aleph}{\mbox{ALEPH}}
\newcommand{\Delphi}{\mbox{DELPHI}}
\newcommand{\Ltre}{\mbox{L3}}
\newcommand{\Opal}{\mbox{OPAL}}

\newcommand{\wwbr}{\mbox{$\mathcal{B}$}}
\newcommand{\Wtoenu}     {\mbox{$W\rightarrow
                                 {e \nu_e}$}}
\newcommand{\Wtomnu}     {\mbox{$W\rightarrow
                                 \mu \nu_\mu$}}
\newcommand{\Wtotnu}     {\mbox{$W\rightarrow
                                 \tau \nu_\tau$}}
\newcommand{\CoM}        {centre-of-mass}

\newcommand{\Wtoqq}      {\mbox{$W\rightarrow
                                 q q$}}
\newcommand{\Htotnu}     {\mbox{$H^\pm\rightarrow
                                 \tau \nu_\tau$}}
\newcommand{\Htoqq}      {\mbox{$H^\pm\rightarrow
                                 q q$}}

\newcommand{\rb}{\frac{\wwbr\mathrm{(\Wtotnu)}}%
{[\wwbr\mathrm{(\Wtoenu)}+\wwbr\mathrm{(\Wtomnu)}]/2}}

\newcommand{\mH}{m_{H^\pm}}
\newcommand{\mW}{m_W}
\newcommand{\MH}{m_H}
\newcommand{\Mh}{m_h}
\newcommand{\MA}{m_A}

\newcommand{\GeV}{\ \mathrm{GeV}}
\newcommand{\tb}{\tan\beta}
\newcommand{\cb}{\cot\beta}
\newcommand{\doublet}[2]{\left(
    \renewcommand{\arraystretch}{1}
    \begin{array}{c}#1\\#2\end{array}\right)}
\newcommand{\dcT}{d \! \cos \! \Theta}
\newcommand{\dcTW}{d \! \cos \! \Theta_{W^-}}

\newcommand{\dtOO}{\dcT \, d \Omega \, d \overline{\Omega}}
\newcommand{\thdm}{2HDM}

\begin{document}

\preprint{hep-ph/0607280}
\preprint{TU-774}

\title{Lepton non-universality at LEP and charged Higgs}

\author{Jae-hyeon Park}
\email{jhpark@tuhep.phys.tohoku.ac.jp}
\affiliation{Department of Physics, Tohoku University, Sendai 980--8578, Japan}

\date{August 18, 2006}

\begin{abstract}
A recent analysis of the LEP data shows an interesting deviation
from lepton universality in $W$ boson decays.
An excess at the level of 2.8~$\sigma$ is found
in the tau mode branching ratio
with respect to the other two modes.
It is suggested that
this seeming lepton non-universality might stem from
pair production of charged Higgs bosons almost degenerate with $W$,
that preferentially decay to heavy fermions.
It is shown that the deviation can be reduced to 1.4~$\sigma$
in two Higgs doublet model~I
without any conflict with the existing direct or indirect constraints.
This conclusion is largely independent of $\tb$,
the ratio of Higgs vacuum expectation values.
This scenario can be tested
at the forthcoming international linear collider.
\end{abstract}

\maketitle

\section{Introduction}

In a non-abelian gauge theory, the gauge interaction
of a fermion is fixed by its representation under the gauge group,
up to an overall gauge coupling.
The charged current interaction in the Standard Model (SM)
is described by an SU(2) group, under which the three
flavors of left-handed leptons transform as doublets.
An obvious consequence is the lepton universality in
the $W$ boson interactions.
This has been tested
in muon decay, leptonic and semi-leptonic tau decays,
and leptonic meson decays (for a review, see~\cite{Pich:1997hj};
updated results can be found in~\cite{Pich:2002bc,Loinaz:2004qc}),
showing no evidence of deviation from it.

In contrast,
recent LEP data on leptonic branching fractions of the
$W$ boson appears to bear a small but intriguing discrepancy
from lepton universality \cite{WpairLEP,Group:2005di}.
An excess of 2.8 standard deviations
is found in the branching ratio of the tau mode
with respect to the other two leptonic modes.

If one takes this number seriously and attempts to give
an account of it with modification to the SM,
a possibility may be to alter the model in such a way that
an (effective) $W$-lepton-neutrino vertex becomes flavor-dependent.
Indeed,
there has been a class of models taking this approach \cite{nonugauge},
the last of which was devised to explain the above discrepancy.
This model has two SU(2) gauge groups.
The first and the second family fermions are charged under one group,
and the third the other.
Mixing of the gauge bosons of different groups can lead to
flavor-dependent lightest $W$ boson couplings to leptons.
Tuning the model parameters, one can accommodate the measured
leptonic branching ratios.
One immediate problem of this idea, however, is that
it is likely to affect the aforementioned muon and tau decay rates
mediated by $W$ exchange, thereby spoiling their agreement with
lepton universality.
Another class of models that can give rise to non-universal charged
current interaction are those involving a low scale seesaw mechanism
\cite{Fujihara:2005uq}, although they
were not conceived for reconciling the leptonic $W$ branching ratios.

As the difficulty with non-universal charged current interaction is
rather evident, this work takes a different approach.
It is speculated that
the apparent excess of tau mode decay rate is in fact
due to pair production of charged Higgs bosons, which
dominantly decay to $\tau\nu$ or $c s$.
The LEP measurements of
the $W$ boson decay rates are performed
by counting the final state fermions.
Thus, if a charged Higgs boson decays,
its decay products may well appear to be coming from a $W$ decay.

To mimic a $W$ boson, a charged Higgs boson should be light enough
to be produced at LEP\@.
Furthermore, its mass should be around $\mW$
to give a meaningful alteration in the number of tau mode events,
since the charged Higgs pair production rate rapidly
decreases as $\mH$ increases.
The following section will elaborate on this.
Throughout this paper,
the charged Higgs mass is assumed to be slightly above the $W$ mass.
This helps to avoid disturbing the $W$ mass measurements
at the pair threshold as well.
Instantly, this somewhat low $\mH$ raises doubts about
its compatibility with the available search results.
Particle search is a model-dependent task, and one should first
specify in which context the charged Higgs is introduced.
Note that this low $\mH$ is hard to be achieved in
the minimal supersymmetric standard model (MSSM)
due to the tree-level relation,
$\mH^2 = \mW^2 + m_A^2$,
in conjunction with the $CP$-odd Higgs mass lower bound,
$m_A > 93\ \mathrm{GeV}$ \cite{mssmhiggslep}.
Thus a remaining natural choice
is a general two Higgs doublet model (\thdm)
which allows for $\mH \approx \mW$.
For this $\mH$, the pair production
cross section of charged Higgs
turns out to be below 1\% of that of $W$, at LEP energies.
This kind of small peak hiding behind a much larger background resonance
is extremely difficult to discriminate \cite{MHeqMW}.
In fact, branching ratio independent charged Higgs mass limit from LEP,
is lower than $\mW$.\footnote{%
A neutral Higgs analogy had been discussed in~\cite{Brown:1990nj}
where the Higgs is assumed to be degenerate with the $Z$ boson.%
}



The two Higgs doublets $H_1$ and $H_2$ in
\thdm\ have eight real degrees of freedom.
Three of them are eaten by the $W$ and $Z$ bosons to be their
longitudinal components.
The remaining five form
the lighter $CP$-even neutral Higgs $h^0$,
the heavier $CP$-even neutral Higgs $H^0$,
the $CP$-odd neutral Higgs $A^0$,
and the charged Higgs $H^\pm$,
in the case of $CP$-conserving Higgs sector.
Henceforth, $CP$-invariance in the Higgs sector
will be assumed.
The five Higgs boson masses in \thdm\ can be changed independently of
one another except that they are subject to
the condition, $\Mh \leq \MH$, following from their definitions.
The Higgs potential has a
large enough number of free parameters for this,
unlike the Higgs sector of MSSM
where many of the parameters are constrained by supersymmetry.

In the presence of more than one Higgs doublets,
flavor changing neural current (FCNC) interactions are in general
mediated by Higgs bosons
at tree level.
In order to avoid this danger, one typically
makes assumptions on the way how Higgs doublets couple to fermions.
This work follows~\cite{Barger:1989fj} in
classifying models with different assumptions.
In Model~I, only one Higgs doublet, say $H_2$,
couples to quarks and charged leptons.
Models~III, IV, and II are obtained from Model~I
by coupling $H_1$, instead of $H_2$,
to the down-type quarks, the charged leptons,
and both, respectively.
These models can be implemented
by adopting a discrete symmetry for example \cite{Gunion:1989we}.
They are summarized in Table~\ref{tab:models}.%
\begin{table}
  \centering
  \renewcommand{\arraystretch}{1.2}
  \begin{tabular}{|c|cc|cc|cc|cc|}
    \hline 
    Models & \multicolumn{2}{c|}{I} & \multicolumn{2}{c|}{II} &
    \multicolumn{2}{c|}{III} & \multicolumn{2}{c|}{IV} \\
    \hline
    &couples to& $A_f$ &couples to& $A_f$ &couples to& $A_f$ &couples to& $A_f$
    \\
    \multirow{2}*{$\doublet{u}{d}$}
    & $H_2$ & $\cb$  & $H_2$ & $\cb$ & $H_2$ & $\cb$ & $H_2$ & $\cb$  \\
    & $H_2$ & $-\cb$ & $H_1$ & $\tb$ & $H_1$ & $\tb$ & $H_2$ & $-\cb$ \\
    \multirow{2}*{$\doublet{\nu}{l}$}
    &       &        &       &       &       &        &       &       \\
    & $H_2$ & $-\cb$ & $H_1$ & $\tb$ & $H_2$ & $-\cb$ & $H_1$ & $\tb$ \\
    \hline
  \end{tabular}
  \caption{For each model,
    the Higgs doublet that couples to each type of fermion
    is shown, and
    the coefficients $A_f$ in the charged
    Higgs Yukawa interaction terms written in~(\ref{eq:int}) are given.
  This table is an excerpt from~\cite{Barger:1989fj}.}
  \label{tab:models}
\end{table}

The interactions between charged Higgs and fermions are given by
\begin{equation}
  \label{eq:int}
  \mathcal{L} = \frac{g}{\sqrt{2} \mW} H^+ [
  V_{ij} m_{u_i} A_u\,\overline{u}_{Ri} d_{Lj}
  + V_{ij} m_{d_j} A_d\,\overline{u}_{Li} d_{Rj}
  + m_l A_l\,\overline{\nu}_L l_R
  ] + \mathrm{h.c.} ,
\end{equation}
where $g$ is the SU(2) gauge coupling,
$V_{ij}$ is an element of the Cabibbo-Kobayashi-Maskawa (CKM) matrix,
and the coefficient $A_f$ for $f = u, d, l$
should be chosen from Table~\ref{tab:models} according to
the given model.
The parameter $\tb$ appearing in this table is defined by
\begin{equation}
  \tb \equiv v_2 / v_1 ,
\end{equation}
with $v_{1,2}$ being the vacuum expectation values of $H_{1,2}$
respectively.
This interaction Lagrangian is important in two respects.
First, it causes the decay $\Htotnu$.
It should have a suitable branching fraction
such that the tau mode excess in $W$ decays
can be attributed to charged Higgs.
Having $\wwbr(\Htotnu)$
within a right range
is also crucial for escaping from
the charged Higgs direct search at LEP\@.
Second, those processes that provide constraints on the model
such as FCNC,
$t \rightarrow H^+ b$, and
$Z \rightarrow b \overline{b}$, are influenced by
the above Yukawa interactions.



The main idea has been outlined up to this point.
The remaining task is then twofold:
to evaluate how much excess of tau production can be
ascribed to the charged Higgs contribution, and
to examine whether or not a variety of experiential
constraints can be satisfied.
This is to be performed for each of the four models listed
in Table~\ref{tab:models} with $\tb$ as a tunable parameter.
Here in advance, it is noted that
only Model~I survives the constraints that
will be discussed in Sec.~\ref{sec:con}.
This leads to $\wwbr(\Htotnu) \simeq 0.7$, which will be used
in the following section.
It will be shown that the lepton non-universality can be
alleviated to a large extent without violating any constraint.
In particular,
the additional tau mode branching fraction
due to charged Higgs, is largely determined 
as a function of $\mH$
independent of $\tb$,
by the direct search
and the $b \rightarrow s \gamma$ constraints.

The paper is organized as follows.
Sec.~\ref{sec:br} contains the main result of this paper,
which reveals that a sizeable fraction of
the observed lepton non-universality
can be resolved by pair production of charged Higgs nearly degenerate with
the $W$ boson.
In Sec.~\ref{sec:con}, it is shown that
the current direct and indirect constraints
are consistent with this mass of charged Higgs.
Sec.~\ref{sec:test} presents discussions on how to test
this scenario at future experiments.
Finally, the conclusion is given in Sec.~\ref{sec:conc}.

\section{Leptonic branching fractions of \boldmath $W$ boson}
\label{sec:br}

The leptonic branching fractions of $W$ boson has been measured
from partial cross sections of $WW \rightarrow 4f$ by the four
experiments at LEP \cite{WpairLEP}
without the assumption of lepton universality.
Part of a table summarizing the result in
the latest report \cite{Group:2005di}
is quoted in Table~\ref{tab:wwbra}\@.%
\begin{table}
\renewcommand{\arraystretch}{1.2}
\begin{center}
\begin{tabular}{|c|c|c|c|}
\hline
Experiment 
         & \wwbr(\Wtoenu) [\%] & \wwbr(\Wtomnu) [\%] & \wwbr(\Wtotnu) [\%] \\
\hline
\Aleph\  & $10.78\pm0.29^*$ & $10.87\pm0.26^*$ & $11.25\pm0.38^*$ \\
\Delphi\ & $10.55\pm0.34^*$ & $10.65\pm0.27^*$ & $11.46\pm0.43^*$ \\
\Ltre\   & $10.78\pm0.32^*$ & $10.03\pm0.31^*$ & $11.89\pm0.45^*$ \\
\Opal\   & $10.40\pm0.35$ & $10.61\pm0.35$ & $11.18\pm0.48$ \\
\hline
LEP      & $10.65\pm0.17$ & $10.59\pm0.15$ & $11.44\pm0.22$ \\
\hline
\end{tabular}
\caption{%
  Summary, copied from \cite{Group:2005di},
  of $W$ branching fractions derived from $W$-pair production
  cross sections measurements up to 207 GeV \CoM\ energy. All results
  are preliminary with the exception of those indicated by $^*$.}
\label{tab:wwbra} 
\end{center}
\end{table}
Interestingly enough, all the four experiments show a tendency
that $\tau \nu_\tau$ mode has a larger branching fraction
than the other two modes, albeit with an error bar not
much shorter than the difference.
A ratio between the tau fraction and the average of electron
and muon fractions is given by
\begin{equation}
\left. \rb \right|_\mathrm{LEP} = 1.077 \pm 0.026 ,
\label{eq:Wbrr}
\end{equation}
under the assumption of partial lepton universality.
The departure from complete lepton universality is at the level of
2.8 standard deviations.
It may be explained away by a statistical fluctuation which
would be gone with sufficient amount of data.
Alternatively, the present work takes an interesting possibility
that this apparent lepton non-universality stems from physics
beyond the minimal Standard Model.

\begin{figure}
  \centering
  \subfigure[]{%
  \begin{picture}(160,116) (90,-70)
    \SetWidth{0.8}
    \SetColor{Black}
    \ArrowLine(115,18)(155,8)
    \ArrowLine(155,8)(155,-32)
    \ArrowLine(155,-32)(115,-42)
    \Photon(155,8)(195,18){2.5}{5}
    \Photon(155,-32)(195,-42){2.5}{5}
    \ArrowLine(195,18)(225,33)
    \ArrowLine(225,3)(195,18)
    \ArrowLine(195,-42)(225,-27)
    \ArrowLine(225,-57)(195,-42)
    \Text(105,18)[]{\normalsize{\Black{$e^-$}}}
    \Text(105,-42)[]{\normalsize{\Black{$e^+$}}}
    \Text(145,-12)[]{\normalsize{\Black{$\nu_e$}}}
    \Text(175,-52)[]{\normalsize{\Black{$W^+$}}}
    \Text(175,26)[]{\normalsize{\Black{$W^-$}}}
    \Text(235,-2)[b]{\normalsize{\Black{$f_2$}}}
    \Text(235,-22)[t]{\normalsize{\Black{$f_3$}}}
    \Text(235,-62)[b]{\normalsize{\Black{$f_4$}}}
    \Text(235,38)[t]{\normalsize{\Black{$f_1$}}}
  \end{picture}}%
\subfigure[]{%
  \begin{picture}(180,116) (95,-72)
    \SetWidth{0.8}
    \SetColor{Black}
    \Photon(150,-14)(190,-14){2.5}{4}
    \Photon(190,-14)(220,16){2.5}{4}
    \Photon(190,-14)(220,-44){2.5}{4}
    \ArrowLine(250,1)(220,16)
    \ArrowLine(220,16)(250,31)
    \ArrowLine(250,-59)(220,-44)
    \ArrowLine(220,-44)(250,-29)
    \Text(260,36)[t]{\normalsize{\Black{$f_1$}}}
    \Text(260,-4)[b]{\normalsize{\Black{$f_2$}}}
    \Text(260,-24)[t]{\normalsize{\Black{$f_3$}}}
    \Text(260,-64)[b]{\normalsize{\Black{$f_4$}}}
    \Text(200,-44)[]{\normalsize{\Black{$W^+$}}}
    \Text(200,14)[]{\normalsize{\Black{$W^-$}}}
    \Text(170,-26)[]{\normalsize{\Black{$\gamma, Z$}}}
    \ArrowLine(150,-14)(120,-44)
    \ArrowLine(120,16)(150,-14)
    \Text(110,-44)[]{\normalsize{\Black{$e^+$}}}
    \Text(110,16)[]{\normalsize{\Black{$e^-$}}}
  \end{picture}}%
\subfigure[]{%
  \begin{picture}(180,116) (95,-72)
    \SetWidth{0.8}
    \SetColor{Black}
    \Photon(150,-14)(190,-14){2.5}{4}
    \ArrowLine(250,1)(220,16)
    \ArrowLine(220,16)(250,31)
    \ArrowLine(250,-59)(220,-44)
    \ArrowLine(220,-44)(250,-29)
    \Text(260,36)[t]{\normalsize{\Black{$f_1$}}}
    \Text(260,-4)[b]{\normalsize{\Black{$f_2$}}}
    \Text(260,-24)[t]{\normalsize{\Black{$f_3$}}}
    \Text(260,-64)[b]{\normalsize{\Black{$f_4$}}}
    \Text(200,-44)[]{\normalsize{\Black{$H^+$}}}
    \Text(200,14)[]{\normalsize{\Black{$H^-$}}}
    \Text(170,-26)[]{\normalsize{\Black{$\gamma, Z$}}}
    \ArrowLine(150,-14)(120,-44)
    \ArrowLine(120,16)(150,-14)
    \Text(110,-44)[]{\normalsize{\Black{$e^+$}}}
    \Text(110,16)[]{\normalsize{\Black{$e^-$}}}
    \DashLine(190,-14)(220,16){5}
    \DashLine(190,-14)(220,-44){5}
  \end{picture}}
  \caption{Tree-level Feynman graphs for $W$-pair production
    within the SM [(a), (b)],
    and charged Higgs pair production in \thdm\ [(c)],
    each followed by subsequent decays to four fermions.}
  \label{fig:4f}
\end{figure}
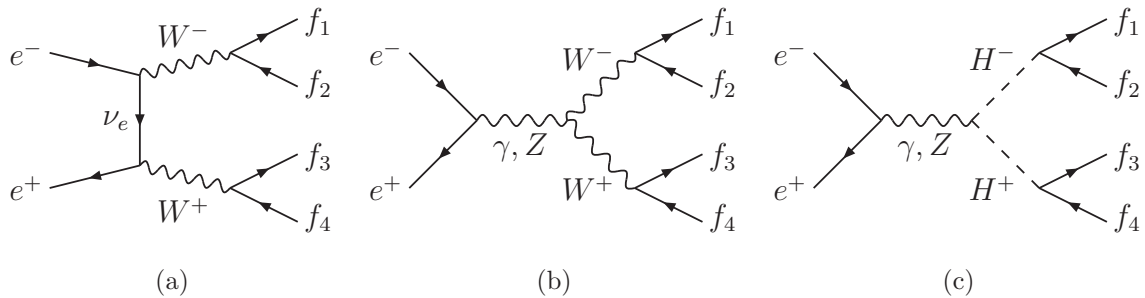

The leptonic $W$ branching fractions were measured from
the $W$-pair production process followed by decays
into four fermions.
In the SM,
the tree-level diagrams for this process are given
by Figs.~\ref{fig:4f}(a) and~\ref{fig:4f}(b).%
The final state fermions can be
$(f_1, f_2) = (e,\overline{\nu}_e), (\mu,\overline{\nu}_\mu),
(\tau,\overline{\nu}_\tau), (d,\overline{u}), (s,\overline{c})$
with equal weight, where $d$ and $s$ quarks should include
CKM mixing, and $(f_4, f_3)$ can be one conjugate of these states.
Now, suppose that there is a charged Higgs boson whose mass
is close to the $W$ boson mass.
Then, the graph in Fig.~\ref{fig:4f}(c) produces four fermions
as well but with different weights,
since a charged Higgs decays preferentially to
$(f_1, f_2) = (\tau,\overline{\nu}_\tau), (s,\overline{c})$.
If one counts the number of final state leptons for each flavor
to measure its branching fraction, the excess of tau final state
caused by charged Higgs decay may appear as a higher
branching fraction of $\Wtotnu$.

It is straightforward to estimate how much difference 
this charged Higgs contamination can make.
The result differs depending on which modes are considered.
The $qql\nu$ modes are most statistically significant.
Comparing only $qq\tau\nu$ and $qq\mu\nu$ for example,
one has the ratio of apparent branching fractions,
\begin{equation}
  \left. \frac{\wwbr(\Wtotnu)}{\wwbr(\Wtomnu)}
    \right|_\mathrm{appar}
  = \frac{\sigma_{WW}^{qq\tau\nu} + \sigma_{HH}^{qq\tau\nu}}%
  {\sigma_{WW}^{qq\mu\nu}}
  = 1 + \frac{\sigma_{HH}}{\sigma_{WW}}
  \frac{\wwbr(\Htotnu)}{\wwbr(\Wtomnu)}
  \frac{\wwbr(\Htoqq)}{\wwbr(\Wtoqq)}
  \approx 1.02 .
  \label{eq:qqln}
\end{equation}
For numerical estimation,
the charged Higgs mass is taken to be 81~GeV and
the center-of-mass energy 200~GeV\@.
For this energy,
the charged Higgs pair production cross section is
$\sigma_{HH} = 0.14\ \mathrm{pb}$ (from the second of \cite{HplusLEP}),
and the $W$ pair production cross section is
$\sigma_{WW} = 17\ \mathrm{pb}$ (from the second of \cite{WpairLEP}).
As for the branching fractions,
$\wwbr(\Wtoqq) = 6/9$, and
$\wwbr(\Wtomnu) = 1/9$ were used, in addition to
$\wwbr(\Htoqq) = 0.3$ and $\wwbr(\Htotnu) = 0.7$
which will be justified in the next section.
One may do a similar calculation with only $\tau\nu\tau\nu$ and
$\mu\nu\mu\nu$ modes to get a higher value, 
\begin{equation}
  \left. \frac{\wwbr(\Wtotnu)}{\wwbr(\Wtomnu)}
    \right|_\mathrm{appar}
    = \sqrt{
      1 + \frac{\sigma_{HH}}{\sigma_{WW}}
      \left(
        \frac{\wwbr(\Htotnu)}{\wwbr(\Wtomnu)}
      \right)^2}
    \approx 1.15,
    \label{eq:lnln}
\end{equation}
although a leptonic channel cross section is much smaller than
a semi-leptonic one, thereby making less significant contribution
to the average.
 From these estimates, one may envisage a possibility that
a few per cent out of
the seeming 8\% deviation of~(\ref{eq:Wbrr}) from unity, is accommodated by
the charged Higgs contribution to four fermion production.

\begin{figure}
  \centering
  \includegraphics[width=8cm]{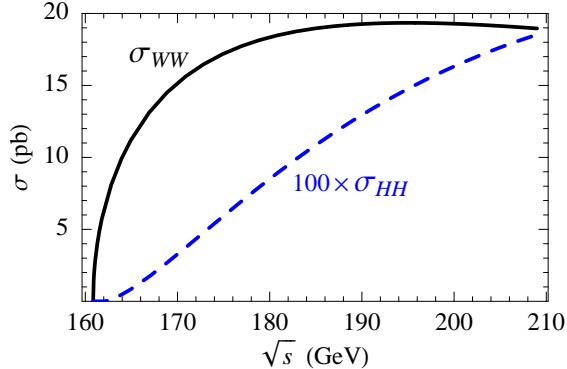}
  \caption{The $W$ boson pair production cross section $\sigma_{WW}$ 
    (the solid curve) and
    the charged Higgs pair production cross section $\sigma_{HH}$
    multiplied by 100 for $\mH = 81\ \mathrm{GeV}$ (the dashed curve),
    from tree-level calculations,
    as functions of the center-of-mass energy available at LEP\@.}
  \label{fig:pair}
\end{figure}

The charged Higgs and $W$-pair production cross sections, however,
are different functions of the center-of-mass energy,
as are shown in Fig.~\ref{fig:pair}.%
Therefore, one should do a more careful analysis
taking into account other factors.
Fortunately, the DELPHI collaboration
(the first of~\cite{WpairLEP}) provides
detailed data
for each of the ten different channels
including the number of selected events,
the background cross section, and the efficiency matrix,
as well as luminosity.
The luminosity and number of events are presented as functions
of $\sqrt{s}$ from 161~GeV to 207~GeV\@.
Using these data, one can perform a maximum likelihood fit.
The likelihood is defined as
\begin{equation}
  L = \prod_{s,i}
  \frac{e^{-\mu_s^i} (\mu_s^i)^{N_s^i}}{(N_s^i)!},
  \label{eq:like}
\end{equation}
where $s$ denotes the squared center-of-mass energy,
$i$ runs over the ten channels,
$jjjj$, $jje\nu$, $jj\mu\nu$, $jj\tau\nu$, $\tau\nu\tau\nu$, $e\nu\tau\nu$,
$\mu\nu\tau\nu$, $e\nu e\nu$, $e\nu\mu\nu$, and $\mu\nu\mu\nu$,
and $N_s^i$ is the number of selected events in channel $i$ at $\sqrt{s}$.
The expected number of events $\mu_s^i$ in channel $i$ at $\sqrt{s}$,
is written as
\begin{equation}
  \mu_s^i = \left(
    \sum_j \epsilon_s^{ij} \sigma_s^j + \sigma_{\mathrm{bg},s}^i
    \right) \mathcal{L}_s ,
\end{equation}
where $\epsilon_s^{ij}$ is the efficiency matrix,
$\sigma_{\mathrm{bg},s}^i$ is the background cross section,
and $\mathcal{L}_s$ is the luminosity.
One can express the channel cross section $\sigma_s^j$ as a function of
$\sigma_{WW,s}$ and the three leptonic $W$ branching fractions,
and use the data in the aforementioned reference for the other variables.
As before, the subscript $s$ attached to a symbol
is a reminder of the relevant center-of-mass energy.

Performing a fit within the SM,
one finds at the maximum of $L$,
\begin{equation}
  \wwbr(\Wtoenu) = 10.79, \quad
  \wwbr(\Wtomnu) = 10.67, \quad
  \wwbr(\Wtotnu) = 11.49.
\end{equation}
These results are slightly different from but
close enough to those by DELPHI
which is reproduced in Table~\ref{tab:wwbra}.
The small variances may have come from the fact that the efficiency matrix
and the background cross sections at $\sqrt{s} = 200\ \mathrm{GeV}$
are used for the other values of $\sqrt{s}$ as well.
Combining the above three numbers leads to
\begin{equation}
  \left. \rb \right|_\text{SM fit} = 1.071 .
\label{eq:smfit}
\end{equation}

Having tested that a reliable estimation can be made
using the available data, one may proceed to do the same fit including
charged Higgs contributions.
This time, the following three channel cross sections should be
modified as
\begin{equation}
  \begin{aligned}
    \sigma_s^{qq\tau\nu} &= \sigma_{WW,s} \cdot
    2 \wwbr(\Wtoqq) \wwbr(\Wtotnu) +
    \sigma_{HH,s} \cdot 2 \wwbr(\Htoqq) \wwbr(\Htotnu) , \\
    \sigma_s^{\tau\nu\tau\nu} &= \sigma_{WW,s} \cdot \wwbr^2(\Wtotnu) 
    + \sigma_{HH,s} \cdot \wwbr^2(\Htotnu), \\
    \sigma_s^{qqqq} &= \sigma_{WW,s} \cdot \wwbr^2(\Wtoqq) 
    + \sigma_{HH,s} \cdot \wwbr^2(\Htoqq),
  \end{aligned}
\end{equation}
where each first term is the usual $W$ boson contribution
and second charged Higgs.
Charged Higgs decays to muon and electron modes are ignored
due to the small Yukawa couplings, and the other modes are assumed to have
$\wwbr(\Htoqq) = 0.3$ and $\wwbr(\Htotnu) = 0.7$ as for~(\ref{eq:qqln})
and (\ref{eq:lnln}).
For $\sigma_{HH,s}$, the tree-level cross section shown in
Fig.~\ref{fig:pair} is used.
With the charged Higgs contributions included,
a likelihood fit results in
\begin{equation}
  \wwbr(\Wtoenu) = 10.84, \quad
  \wwbr(\Wtomnu) = 10.73, \quad
  \wwbr(\Wtotnu) = 11.12.
\end{equation}
Recall that these are not the apparent branching fractions of
the $W$ boson,
but the real ones
excluding contaminations from charged Higgs decays.
Thus, their ideal values should coincide with one another.
The tau mode excess, then, decreases to
\begin{equation}
  \left. \rb \right|_\text{2HDM fit} = 1.031.
  \label{eq:r}
\end{equation}
Comparing this and (\ref{eq:smfit}),
one can notice that the deviation from unity has been diminished by
4\%, a value in between the estimates
from~(\ref{eq:qqln}) and (\ref{eq:lnln}).

It is reasonable to guess that approximately the same amount of
reduction can be made for each of the four experiments.
This would reduce the overall ratio to
\begin{equation}
\left. \rb \right|_\mathrm{LEP,2HDM} \simeq 1.037 \pm 0.026 ,
\end{equation}
of which the departure from unity is $1.4\ \sigma$.
This is a noticeable, if not complete,
amelioration from the original $2.8\ \sigma$
in~(\ref{eq:Wbrr}).

\begin{figure}
  \centering
  \includegraphics[width=8cm]{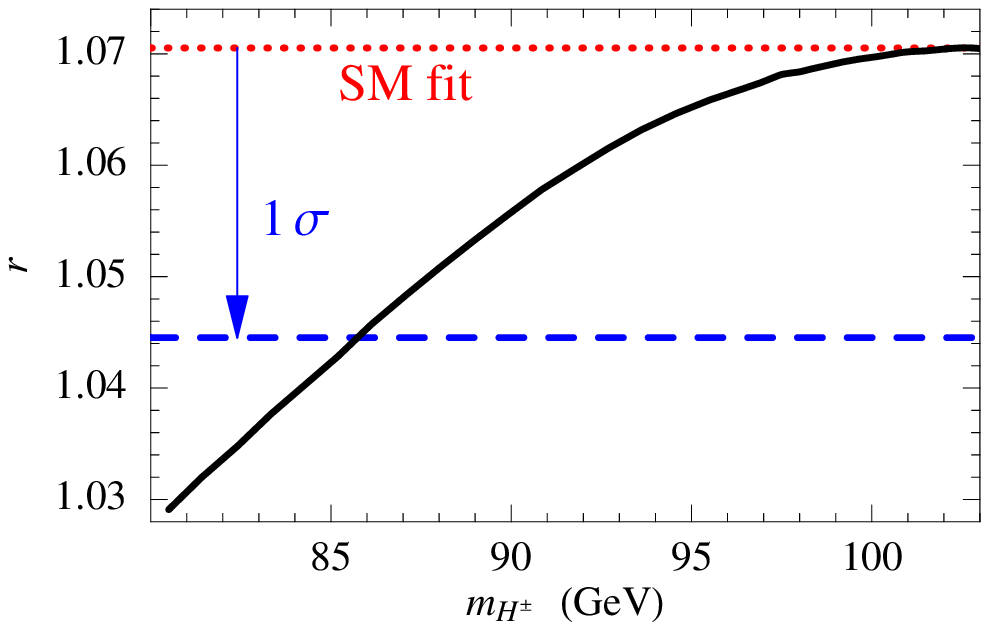}
  \caption{Fit result of
    $r \equiv 2\wwbr(\Wtotnu) / [\wwbr(\Wtoenu)+\wwbr(\Wtomnu)]$
    using the DELPHI data,
    as a function of $\mH$.
    The Model~I branching fractions,
    $\wwbr(\Htoqq) = 0.3$ and $\wwbr(\Htotnu) = 0.7$, are used.
  The length of the vertical arrow is one standard deviation
  from~(\ref{eq:Wbrr}).}
  \label{fig:MHc}
\end{figure}

Remember that this prediction is a function of $\mH$ and $\wwbr(\Htotnu)$,
and $\mH$ has been assumed to be 81~GeV so far.
A remark is in order on how the result depends on these parameters.
Once one assumes that the charged Higgs mass is close to the $W$ boson mass
and imposes the charged Higgs direct search limit and
$b \rightarrow s \gamma$ constraint,
$\wwbr(\Htotnu)$ is almost uniquely determined,
as will be discussed in the next section.
The remaining dependence on $\mH$ is shown in Fig.~\ref{fig:MHc}.%
The vertical axis, labeled $r$, is the ratio in~(\ref{eq:r})
from the fit using the DELPHI data for a given value of $\mH$.
The ratio worsens as the charged Higgs gets heavier, and eventually
for $\mH = 103\ \mathrm{GeV}$, coincides with
the SM fit in~(\ref{eq:smfit}), marked by the dotted horizontal line.
Obviously, a light charged Higgs is favored for lepton-universality.
To reduce the ratio by at least 1-$\sigma$ ($= 0.026$), one should require that
\begin{equation}
  \mH < 85.7 \ \mathrm{GeV}.
  \label{eq:maxMHc}
\end{equation}
It can be checked that this inequality holds for any value of $\wwbr(\Htotnu)$.
This justifies the assumption of $\mH \approx \mW$ made above.

\section{Constraints from data}
\label{sec:con}


\subsection{\boldmath $B \rightarrow X_s \gamma$}

One of the most stringent lower limits on the charged Higgs mass
has been given by the process $b \rightarrow s \gamma$.
One will see that
this constraint, combined with the direct search limit from LEP,
almost determines the type of \thdm\ that can be used
for the present purpose.

In order to notice different behaviors among models,
leading logarithmic approximation should be enough here.
The weak effective Hamiltonian for the $B \rightarrow X_s \gamma$
mode reads
\begin{equation}
  \mathcal{H}_\mathrm{eff} = - \frac{4 G_F}{\sqrt{2}}
  V^*_{ts} V_{tb} \sum_{i=1 \ldots 6,7\gamma,8g} C_i(\mu) Q_i(\mu)
  + \mathrm{h.c.},
\end{equation}
where operators with non-vanishing Wilson coefficients at the matching
scale are
\begin{equation}
  \begin{aligned}
    Q_2 &= \overline{s}_L \gamma_\mu c_L \ \overline{c}_L \gamma^\mu b_L ,\\
    Q_{7\gamma} &= \frac{e}{16\pi^2}\,
    m_b\,\overline{s}_L \sigma_{\mu\nu} F^{\mu\nu} b_R ,\\
    Q_{8g} &= \frac{g_s}{16\pi^2}\,
    m_b\,\overline{s}_L \sigma_{\mu\nu} G^{\mu\nu} b_R .
  \end{aligned}
\end{equation}
In the SM,
the Wilson coefficients
at the matching scale $\mu_W$ are given by
\begin{align}
    C^\mathrm{SM}_2 (\mu_W) &= 1, &
    C^\mathrm{SM}_{7\gamma,8g} (\mu_W) &= F^{(1)}_{7,8} (x) , \\
\intertext{and in \thdm, there are additional charged Higgs contributions
\cite{Ciuchini:1997xe},}
    C^{H^\pm}_2 (\mu_W) &= 0, &
    C^{H^\pm}_{7\gamma,8g} (\mu_W) &= \frac{A_u^2}{3} F^{(1)}_{7,8} (y)
    + A_u A_d F^{(2)}_{7,8} (y) ,
\end{align}
with the notations,
$x \equiv m_t^2 / \mW^2$ and $y \equiv m_t^2/\mH^2$.
Definitions of $F^{(1,2)}_{7,8}$ can be found in~\cite{Ciuchini:1997xe}.
After performing renormalization group running of these
Wilson coefficients down to the $m_b$ scale \cite{Kagan:1998bh},
one has the ratio of branching fractions in \thdm\ and in the
SM,
\begin{equation}
  \frac{\wwbr(B \rightarrow X_s \gamma)}%
  {\wwbr_\mathrm{SM} (B \rightarrow X_s \gamma)} =
  \left|
  \frac{C^\mathrm{SM}_{7\gamma} (m_b) + C^{H^\pm}_{7\gamma} (m_b)}%
  {C^\mathrm{SM}_{7\gamma} (m_b)}
  \right|^2
  = \left| 1 + 0.69 A_u A_d + 0.14 A_u^2 \right|^2 ,
\end{equation}
where it is assumed that $\mH = 85.7\ \mathrm{GeV}$ from~(\ref{eq:maxMHc}).

 From Table~\ref{tab:models},
it is obvious that Models~II and III
lead to at least 180\% excess of
$\wwbr(B \rightarrow X_s \gamma)$ with respect to the SM
result for any value of $\tb$ because of the term
proportional to $A_u A_d$.
This is in gross conflict with the data, and thus these two models
are excluded from consideration.
However, for the remaining two models, sizes of both $A_u$ and $A_d$
are inversely proportional to $\tb$.
Indeed, the difference in branching fraction made by charged Higgs
contributions is reduced below the current experimental error
\cite{hfag2006}
provided that
\begin{equation}
  \tb \gtrsim 4 .
\end{equation}
This is the case even for $\mH = 81\ \mathrm{GeV}$.
Consequently,
one can conclude that Models~I and IV are consistent with
$\wwbr(B \rightarrow X_s \gamma)$
for moderately high $\tb$ \cite{Barger:1989fj,Ciuchini:1997xe}.
This behavior of charged Higgs decoupling from quarks
for high $\tb$, will
be crucial for evading low energy constraints discussed later.


\subsection{Direct searches}
\label{sec:direct}

\paragraph{$H^+ H^-$ pair production at LEP}

\begin{figure}
  \centering
  \includegraphics[width=8cm]{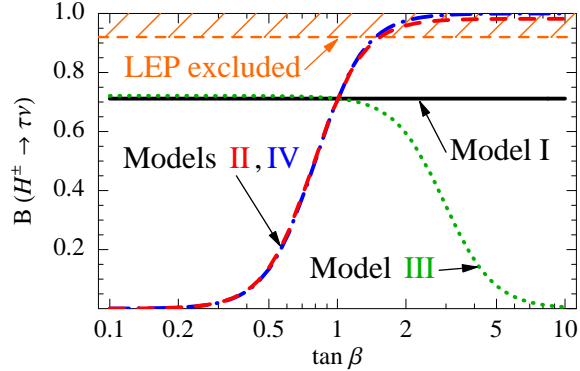}
  \caption{The branching fraction $\wwbr(\Htotnu)$ as a function
    of $\tb$ for each of the Models~I--IV, with the parameters,
    $m_c = 0.67\ \mathrm{GeV}$, $m_s = 0.07\ \mathrm{GeV}$,
    $V_{cs} = 1$, and
    $m_\tau = 1.777\ \mathrm{GeV}$.
    The running quark masses are from~\cite{Caravaglios:2002br}.
    The hatched range of $\wwbr(\Htotnu)$ is excluded
    by the charged Higgs direct search at LEP for $\mH = 85.7\ \mathrm{GeV}$.
    This is an update of a plot in~\cite{Barger:1989fj}.}
  \label{fig:Hbr}
\end{figure}

The four experiments at LEP have performed direct searches for
charged Higgs by pair production thereof \cite{HplusLEP}, described by
Fig.~\ref{fig:4f}(c).
The resulting lower bound on the charged Higgs mass varies
rather significantly depending on
the branching fraction $\wwbr(\Htotnu)$.
Examining the exclusion plots, one can notice
that $\mH = 85.7\ \mathrm{GeV}$
is not excluded by the data at 95\% confidence level provided that
\begin{equation}
\wwbr (\Htotnu) < 0.92 .
\label{eq:brconst}
\end{equation}
The branching fraction for each model as a function of $\tb$
is presented in Fig.~\ref{fig:Hbr}.%
Among the four models shown in the plot,
only Models~I and IV were allowed by
$b \rightarrow s \gamma$ for $\tb \gtrsim 4$
in the previous subsection.
The figure shows that
Model~IV leads to $\wwbr (\Htotnu) \gtrsim 0.99$ for this range of $\tb$.
As a result, the only remaining possibility is Model~I in which
\begin{equation}
  \wwbr(\Htotnu) \simeq 0.7.
  \label{eq:Hbr1}
\end{equation}
For this value, the current charged Higgs mass lower bound is
\begin{equation}
  \mH > 80.7 \GeV
\end{equation}
at 95\% confidence level.
This is the reason why 81~GeV is used in this paper
as a representative value of $\mH$ that maximizes charged Higgs effects.

Note that the branching fraction in~(\ref{eq:Hbr1}) is independent of $\tb$,
a property not shared by the other three models.
This makes it easier to
avoid a plethora of low energy constraints involving
the charged Higgs Yukawa couplings to fermions
by raising $\tb$.
As $\tb$ increases, $\wwbr(\Htotnu)$ does not grow, so that
the requirement~(\ref{eq:brconst}) remains obeyed.
Neither does it decrease, so that
sufficient tau production can be achieved.
Nevertheless, one cannot raise $\tb$ all the way to infinity
since the lifetime of charged Higgs grows like $\tan^2\beta$.
If its decay vertex is far away from
the place of $e^+ e^-$ collision, it may not look like a $W$ boson
decay.
For $\tb = 100$, the decay length is about 7~nm, which is negligible
compared to the detector dimensions.
Therefore, one may forget about it for a reasonable value
of $\tb$.

The main background hindering the charged Higgs searches
is due to $W$-pair production
given by Figs.~\ref{fig:4f}(a) and~\ref{fig:4f}(b).
Because of this,
these searches cannot be very sensitive for $\mH$ around $\mW$.
In a sense, the present work is exploiting this fact
the other way around
in order to use charged Higgs as a source of background of the $W$ boson
production.

\paragraph{$W$-pair production cross section at LEP}

If the four fermion final state from a charge Higgs pair is confused
with that from a $W$-pair, then
the $W$-pair production cross section will appear to have an excess.
However, this excess is smaller than the current error.
The length of error bar of $\sigma_{WW}$ measured at LEP
ranges between $0.21\ \mathrm{pb}$ and $0.7\ \mathrm{pb}$
\cite{Group:2005di,WpairLEP}, depending on $\sqrt{s}$.
This is larger than $\sigma_{HH}$ plotted in Fig.~\ref{fig:pair},
from which one can read that
$\sigma_{HH} \lesssim 0.2\ \mathrm{pb}$ for the center-of-mass
energies available at LEP\@.

\paragraph{Angular distribution of $W$-pair production at LEP}

The angular distribution of $W$-pair production
can place a constraint on the charged Higgs contamination
as the two have different angle dependences.
The LEP measurements of the differential cross section
$d \sigma_{WW} / \dcTW$, though, are selecting only
$qqe\nu$ and $qq\mu\nu$ final states for this purpose since
from a jet, it is hard to tell the charge of the $W$ boson
which decayed into the jet \cite{Group:2005di,WpairLEP}.
A charged Higgs seldom decays to $e\nu$ or $\mu\nu$, and hence
its contribution to the observed $d \sigma_{WW} / \dcTW$
is negligible.

\paragraph{Anomalous triple-gauge-boson couplings of $W$ at LEP}

Although the analysis of $d \sigma_{WW} / \dcTW$
discards $qqqq$, $qq\tau\nu$, and $\tau\nu\tau\nu$ events,
triple-gauge-boson coupling measurements make use of them \cite{tgcdata}.
Thus, one should check whether or not the charged Higgs contribution
affects them excessively.
It is convenient to follow the convention from~\cite{Hagiwara:1986vm},
as in the above references for experimental data.
The latest analysis by the L3 collaboration
(the second of \cite{tgcdata}) reports
the six parameters given in Table~\ref{tab:tgc}.%
\begin{table}
\renewcommand{\arraystretch}{1.2}
  \centering
  \begin{tabular}{|c|l|l|l|}
    \hline
    &\multicolumn{1}{c|}{$g^Z_1$}&\multicolumn{1}{c|}{$\kappa_\gamma$}&
    \multicolumn{1}{c|}{$\lambda_\gamma$} \\
    \hline
    SM & 1.0 & 1.0 & $\phantom{-}0.0$ \\
    L3 & $0.966^{+0.034}_{-0.032}\pm 0.015$ &
    $0.910^{+0.074}_{-0.066}\pm 0.039$ & $-0.024^{+0.035}_{-0.033}\pm 0.017$ \\
    2HDM & $1.006$ & $0.956$ & $-0.000$ \\
    \hline
    &\multicolumn{1}{c|}{$g^Z_5$}&\multicolumn{1}{c|}{$\kappa_Z$}&
    \multicolumn{1}{c|}{$\lambda_Z$} \\
    \hline
    SM & 0.0 & 1.0 & $\phantom{-}0.0$ \\
    L3 & $0.00\pm 0.13\pm 0.05$ &
    $0.924^{+0.059}_{-0.056}\pm 0.024$ & $-0.088^{+0.060}_{-0.057}\pm 0.023$ \\
    2HDM & $0.000$ & $1.009$ & $\phantom{-}0.009$ \\
    \hline
  \end{tabular}
  \caption{The triple-gauge-boson couplings
    $g^Z_1$, $\kappa_\gamma$, $\lambda_\gamma$,
    $g^Z_5$, $\kappa_Z$, and $\lambda_Z$,
    given in the SM,
    and their one-parameter fit results from the L3 experiment
    (the second of \cite{tgcdata}).
    Each row labeled \thdm\ shows
    the fit results including four fermion production due to
    charged Higgs pair for $\mH = 81\GeV$.
  }
  \label{tab:tgc}
\end{table}
The results are for the $W$-pair data.

One may estimate
the influence of charged Higgs pair on these parameters
using an ideal experiment.
Imagine an angular distribution measurement with 100\% efficiency.
For infinite integrated luminosity, one could replace
the variables in the likelihood function of~(\ref{eq:like}),
with the differential cross sections, as
\begin{equation}
  \begin{aligned}
  N^i_a =
      \frac{d\sigma^i_{WW}[(\Theta, \Omega, \overline{\Omega})_a]}{\dtOO} +
      \frac{1}{(4\pi)^2} \frac{d\sigma^i_{HH}(\Theta_a)}{\dcT} ,
      \quad
  \mu^i_a (\Psi) =
  \frac{d\sigma^i_{WW}[(\Theta, \Omega, \overline{\Omega})_a ;\Psi]}{\dtOO} .
  \end{aligned}
\end{equation}
Here, $N^i_a$ corresponds to the observed number of events
in which the $W$ or charged Higgs pair decays into
a set $i$ of four fermions for a set $a$ of angles.
The angle $\Theta$ is between $e^-$ and $W^-$ or $H^-$ momentum directions
in the $e^+ e^-$ center-of-mass frame, and $\Omega$ ($\overline{\Omega}$)
is the direction of $f_1$ ($f_4$) momentum in the $W^-$ ($W^+$) rest
frame.
The expected number $\mu^i_a(\Psi)$ is expressed as a function
of a given set of couplings $\Psi$.
By maximizing the likelihood function, one obtains the values
in the table labeled 2HDM,
for which the charged Higgs mass is taken to be 81~GeV\@.
The difference of each coupling from its SM value is
much smaller than or comparable to the error.
Therefore, one can conclude that $\mH \approx \mW$ is compatible
with the triple-gauge-boson coupling measurements.
This should not be a surprise given that this mass
of charged Higgs is allowed by the direct search discussed above
which already incorporates angular distribution information of
the final state fermions.

\paragraph{$W$ mass and width measurements at LEP}

If the charged Higgs is not exactly degenerate with the $W$ boson,
its invariant mass distribution will be distorted, thereby disturbing
the $W$ mass and width measurements from direct reconstruction.
This effect, though, is negligible due to the small
production rate of charged Higgs pair.
Using the same method as for checking
the anomalous triple-gauge-boson couplings,
one can estimate the shifts made by additional charged Higgs peak,
in the $W$ mass and width determined from Breit-Wigner distribution fit.
The shifts are smaller than the current errors \cite{Group:2005di},
for $\mH$ between 81~GeV and 85~GeV\@.

\paragraph{$t \rightarrow H^+ b$ at CDF}

Since the charged Higgs boson mass in consideration is smaller than
$m_t - m_b$, it can be produced in the top quark decay.
A recent charged Higgs branching
ratio independent analysis
at CDF \cite{Abulencia:2005jd}
indicates that the mass of $\mH = 81\GeV$
is consistent with the data provided
\begin{equation}
  \wwbr (t \rightarrow H^{\pm} b) \lesssim 0.75 .
\end{equation}
For Model~I, one can translate the above bound into the condition,
$\tb \gtrsim 0.5$,
assuming that the top quark decays only to $W^+ b$ or $H^+ b$.
If a more sophisticated analysis tightens the upper bound on
the branching fraction, $\tb$ may need to be increased by
a factor of a few.


\subsection{Indirect constraints}

\paragraph{$S$, $T$, and $U$}

Extended Higgs sector leads to additional oblique corrections
to gauge boson propagators,
which can be characterized by three parameters, $S$, $T$, and $U$
\cite{peskintakeuchi}.
Since these corrections arise from gauge couplings of the Higgs bosons,
constraints on the $S$, $T$, and $U$ parameters are not satisfied
simply by increasing $\tb$,
unlike constraints on flavor changing processes.
In order to estimate the additional effects,
one can define $\delta S$, $\delta T$, and $\delta U$ relative to
some reference Standard Model, following~\cite{Haber:1992cn}.
Their expressions in \thdm\ are available in~\cite{Haber:1992cn,Haber:1999zh}.
They depend on the difference between $\beta$ and
the $CP$-even neutral Higgs mixing angle $\alpha$, and
the Higgs masses, $\mH$, $\Mh$, $\MH$, and $\MA$.

\begin{figure}
  \centering
  \includegraphics[width=8cm]{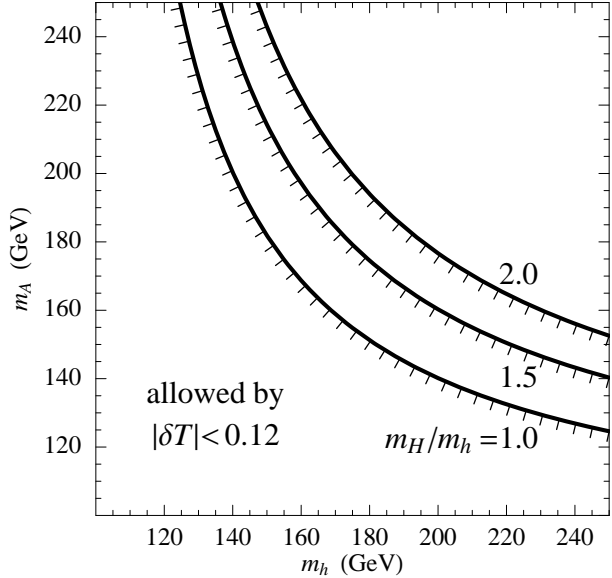}
  \caption{Constraints on $\Mh$ and $\MA$
    from $\delta T$,
    with the ratio $\MH/\Mh$ fixed at 1.0, 1.5, and 2.0,
    respectively.
    The charged Higgs mass is assumed to be 81~GeV\@.
    For a given $\MH/\Mh$, some value of $\alpha - \beta$ leads to
    $|\delta T| < 0.12$, the current 1-$\sigma$ range,
    in the lower left part of the corresponding curve.}
  \label{fig:tcontour}
\end{figure}

The fit result of $T$ using the electroweak data
is that $T = -0.17 \pm 0.12$ for 117~GeV of Higgs mass
\cite{Eidelman:2004wy}.
This error bar is used in the requirement $|\delta T| < 0.12$,
which leaves the allowed region for $\mH = 81\GeV$
on the $(\Mh, \MA)$ plane 
displayed in Fig.~\ref{fig:tcontour}.%
At each point on the plane, $\MH$ is determined by a given value of
$\MH/\Mh$, and $\alpha - \beta$ is chosen such that it minimizes $|\delta T|$.
This optimal value of $\delta T$ is used to draw the three exclusion plots,
for $\MH/\Mh$ fixed at 1.0, 1.5, and 2.0, respectively.
There appears a tendency that $\delta T$ increases as
neutral Higgs bosons get heavier.
It is known that the size of correction to $T$ grows with
the mass split between charged Higgs and neutral Higgs bosons
\cite{Tmasssplit}.
Nevertheless, it is clear that there is plenty of allowed
parameter space with light enough $h^0$ or $A^0$.


The constraint from $\delta S$ is not so severe as the one from
$\delta T$. 
The influence on $\delta U$ is completely negligible compared
to those on $T$ and $S$.

\paragraph{FCNC processes and $CP$ violation}

Flavor and $CP$ violating processes
supply important constraints on \thdm.
As was the case for $B \rightarrow X_s \gamma$, however,
charged Higgs contributions to these processes
get suppressed as $\tb$ increases, even if $\mH \approx \mW$.

The first FCNC constraint to check
is neutral meson mixing. 
Using the expressions in~\cite{Barger:1989fj},
one finds that
for $\tb \gtrsim 4$, the value of $\Delta M_{B_d}$
approaches the SM result within 10\%.
The latter agrees
with the data up to hadronic uncertainties of about 20\%.
Therefore, the $\Delta M_{B_d}$ constraint can be obeyed. 
As for $B_s$--$\overline{B_s}$ mixing,
it is usual to consider
the ratio $\Delta M_{B_s} / \Delta M_{B_d}$ of mass splittings
of $B_s$ and $B^0$ mesons as hadronic uncertainties can be reduces
in this way.
Since the top quark contribution dominates
in the box graphs for both types of mixings,
the ratio is the same as in the SM, which
is in agreement with the recent measurements \cite{bsbsbar}.
In this model, the CKM matrix is the only source of flavor and $CP$ violation,
and hence $\sin 2\beta$ measured from $B \rightarrow J/\psi K$
is unchanged either.
In the case of $\Delta M_K$, its theoretical prediction is
highly uncertain due to long distance effects, and
the constraint should be safe for $\tb$ consistent with
$B^0$--$\overline{B^0}$ mixing.

It can also be expected that increasing $\tb$
to an appropriate extent can make $CP$ violation observables
such as $\epsilon_K$ \cite{Barger:1989fj,Grossman:1994jb,Buchalla:1990fu},
and $\epsilon'/\epsilon_K$ \cite{Barger:1989fj,Buchalla:1990fu}
compatible with experiments.
The same conclusion holds for
other flavor changing processes such as
$K_L \rightarrow \mu^+ \mu^-$ \cite{Barger:1989fj}
and $K^+ \rightarrow \pi^+ \nu \overline{\nu}$
\cite{Barger:1989fj,Grossman:1994jb}.
Recently Belle announced an evidence of purely leptonic decay
$B^- \rightarrow \tau \overline{\nu}$ \cite{Ikado:2006un},
which can be mediated by a charged Higgs in addition to a $W^-$ boson
\cite{Hou:1992sy}.
For $\tb \gtrsim 1$, the branching fraction of this decay
coincides with its SM value within 1\%,
while the data still has around 30\% of error.

\paragraph{$A_b$ and $R_b$}

Constraints on \thdm\ from $A_b$ and $R_b$ in the
$Z \rightarrow b \overline{b}$ process have been studied
in~\cite{Haber:1999zh}.
It was shown that Model~I with
$\mH = 81\GeV$ for $\tb \gtrsim 2$
was consistent with
the then measured value, $R_b = 0.21642 \pm 0.00073$.
This is because charged Higgs decouples from fermions
for high $\tb$.
The error has been reduced by 10\% by now \cite{Eidelman:2004wy},
but the conclusion
should not change very much.
The data on $A_b$ is not as constraining as $R_b$.

\paragraph{$\mu$, $\tau$, $\pi$, and $K$ decay}

Universality of $W$ boson couplings to lepton charged currents
has been thoroughly tested in leptonic decays of muon and tau.
In contrast to the apparent disagreement observed at LEP,
these decay modes show a perfect agreement with lepton universality
with much smaller errors.
A summary of this fact using recent data can be found in~\cite{Loinaz:2004qc},
which states that
\begin{align}
  (g_\mu / g_e)_\tau = 0.9999 \pm 0.0021, \\
  (g_\tau / g_e)_{\tau\mu} = 1.0004 \pm 0.0022, \label{eq:taumuratio}
\end{align}
where $g_{e,\mu,\tau}$ are $W$ boson couplings to
electron, muon, and tau currents, respectively.
The first ratio was derived from tau decay rates, and the second
tau and muon decay rates, as indicated by the subscript outside
each pair of parentheses.
This observation will be maintained if
charged Higgs exchange does not make more than 0.4\% of difference
in a decay rate.

In Model~I, the decay rate of
$\tau \rightarrow \mu \nu \nu$ is roughly reduced
by a factor $(1 - 2 \cot^2\beta\, m_\mu^2 / \mH^2)$ \cite{taudecay}.
Requiring that this decay rate does not change more than 0.4\%,
for $\mH = 81\GeV$,
leads to $\tb \gtrsim 0.03$, and
$\tau \rightarrow \mu \nu \nu$ is virtually unaffected
for values of $\tb$ compatible with $b \rightarrow s \gamma$.
Changes in other modes are even less significant due to smaller lepton
Yukawa couplings.
As a consequence,
leptonic tau and muon decay constraints can remain satisfied,
which is also true of
pion, kaon, and semi-leptonic tau decays for the same reason.

It should be emphasized that this property of the present approach
is different than a class of models which have
flavor-dependent effective $W$ boson couplings to charged currents.
As mentioned earlier, a model
has been proposed \cite{nonugauge} that is claimed to account for
the observed differences in leptonic $W$ decay rates.
In this model, the effective $W$-$\tau$-$\nu_\tau$ coupling
is enhanced by 3.4\% relative to that of
$W$-$\mu$-$\nu_\mu$ or $W$-$e$-$\nu_e$
through mixing of different SU(2) gauge bosons that selectively couple
to different flavors.
Although this renders the $W$ branching fractions 
consistent with~(\ref{eq:Wbrr}),
it decreases the ratio of
$\Gamma(\tau \rightarrow \mu \nu \nu) /
\Gamma(\mu \rightarrow e \nu \nu)$
by 7\% from its SM value,
thereby conflicting with~(\ref{eq:taumuratio}).




%

\section{Test at future experiments}
\label{sec:test}

The defining character of this scenario is
the existence of a charged Higgs boson, with its mass close to $\mW$,
that couples very weakly to fermions.
Therefore, its test can be reduced to charged Higgs search.
The preferred way is
pair production via gauge interactions since
charged Higgs Yukawa couplings may be tiny for high $\tb$.
Although the LEP experiments
still allow the $\mH$ considered in this work,
a similar analysis should be able to rule out
$\mH \approx \mW$ with higher luminosity at a future $e^+ e^-$ machine
such as the international linear collider (ILC),
in spite of the large background from $W$-pairs.

Apart from the higher luminosity, ILC offers an interesting possibility
of utilizing polarized electron and positron beams.
The pair production cross sections
of charged Higgs and $W$ are shown in Table~\ref{tab:ilc}.%
\begin{table}
  \centering
  \begin{tabular}{|c|c|c|c|}
    \hline
    $e^-$/$e^+$ polarization &$\sigma_{HH}$ [pb]&$\sigma_{WW}$ [pb]&
    $\sigma_{HH}/\sigma_{WW}$ [\%]\\
    \hline
    \phantom{8}0\%/\phantom{8}0\% & 0.10 & 7.13 & \phantom{1}1.4 \\
              80\%/\phantom{8}0\% & 0.05 & 1.47 & \phantom{1}3.3 \\
              90\%/\phantom{8}0\% & 0.04 & 0.76 & \phantom{1}5.4 \\
              80\%/60\%           & 0.06 & 0.65 & \phantom{1}8.7 \\
              90\%/60\%           & 0.06 & 0.37 &           15.0 \\
    \hline
  \end{tabular}
  \caption{Pair production cross sections of charged Higgs and $W$,
    and their ratio,
  at $\sqrt{s} = 500\ \mathrm{GeV}$ for different
  right-handed electron and left-handed positron beam polarizations.
  The charged Higgs mass is assumed to be 81~GeV\@.}
  \label{tab:ilc}
\end{table}
They are tree-level values at $\sqrt{s} = 500\ \mathrm{GeV}$
for right-handed polarized electron and
left-handed polarized positron beams.
The charged Higgs mass is assumed to be 81~GeV\@.
One can notice that the beam polarizations can improve
the signal-to-background ratio.

An indirect test is to measure
the decays of pair-produced $W$ bosons
into different quark flavors.
This may be considered as a background suppression method
in charged Higgs search.
If the charged Higgs
is the reason for the apparent excess in $\wwbr(\Wtotnu)$,
it should also cause the same kind of excess in
$\wwbr(W \rightarrow cs)$.

Rejection of the $W$ boson background in an hadronic environment
such as LHC should not be as easy as at ILC.



\section{Conclusion}
\label{sec:conc}

A resolution is proposed
of the possible lepton non-universality observed at the $W$-pair production
experiments at LEP\@.
Introduction of a charged Higgs boson
with $\mH = 81\GeV$,
within the framework of \thdm,
could reduce the 2.8~$\sigma$ of deviation down to 1.4~$\sigma$.
In this way, the excessive tau mode decay rate is attributed to
the pair production of charged Higgses, which
decay preferentially to $\tau\nu$
among the three lepton flavors.

This can be achieved without conflict with the existing
direct or indirect constraints.
In particular, charged Higgs direct search at LEP
in combination with $b \rightarrow s \gamma$
singles out one viable type of \thdm\
out of the four that are free of tree-level
FCNC interactions.
This, in turn, determines what fraction of the tau production
can be ascribed to charged Higgs,
without $\tb$ dependence.
Another point to note is that
this approach does not spoil other lepton universality tests
from muon, tau, and meson decays.

This scenario can be tested at ILC by
charged Higgs direct search.

\begin{acknowledgments}
The author thanks
Andrew G.~Akeroyd,
P.~Ko,
Yukinari Sumino,
Reisaburo Tanaka, and
Masahiro Yamaguchi
for helpful discussions.
This work was supported by the JSPS postdoctoral fellowship program
for foreign researchers and the accompanying grand-in-aid no.\ 17.05302.
\end{acknowledgments}

\end{document}